\documentclass[11pt]{article}
\usepackage{graphicx}

\usepackage{cite}
\usepackage{citesupernumber}
\usepackage{nature}
\usepackage{doublespace}
\title{Navigating the Small World Web by Textual Cues\footnote{This
work is funded in part by NSF CAREER Grant No. IIS-0133124.}}

\author{Filippo Menczer \\
Department of Management Sciences \\
The University of Iowa \\
Iowa City, IA 52242 \\
Phone: (319) 335-0884 \\
Fax: (319) 335-0297 \\
\texttt{filippo-menczer@uiowa.edu}
}

\begin{document}

%\date{}

\begin{singlespace}
\maketitle
\end{singlespace}

\textbf{Can a Web crawler efficiently locate an unknown relevant page?
While this question is receiving much empirical attention due to its 
considerable commercial value in the search engine 
community\cite{Cho98,Chakrabarti99,Menczer00,Menczer01}, theoretical 
efforts to bound the performance of focused navigation have only exploited 
the link structure of the Web graph, neglecting other 
features\cite{Kleinberg01,Adamic01,Kim02}. 
Here I investigate the connection between linkage and a content-induced 
topology of Web pages, suggesting that efficient paths can be discovered 
by decentralized navigation algorithms based on textual cues.}

\newpage

Topic driven crawlers\cite{Cho98,Chakrabarti99,Menczer00,Menczer01} 
are increasingly seen as a way to address the 
scalability limitations of universal search engines, by distributing 
the crawling process across users, queries, or even client computers.  
The context available to such crawlers can guide the navigation of links
with the goal of efficiently locating highly relevant target pages.
Given the need to find unknown target pages, 
we are only interested in decentralized crawling algorithms, which can only use 
information available locally about a page and its neighborhood.
Starting from some source Web page, we aim to visit a target page 
by navigating a path of length $\ell \ll N$ where $\ell$ is the number of
pages visited along the path and $N$ is the total number of pages. 

Since the Web is a small-world network\cite{Albert99} we know that the diameter
(at least for the largest connected component\cite{Broder00,Kleinberg01}) scales 
logarithmically with N, therefore some short path exists between the source and
target nodes such that $\ell \sim \log N$. Can a crawler navigate such a short path?
If the only local information available is about the hypertext link degree of each node 
and its neighbors, then simple greedy algorithms that always pick the neighbor with
highest degree lead to paths where the number of links traversed $\ell'$ scales 
sublinearly\cite{Adamic01} ($\ell' \sim N^{\beta}, \beta < 1$) or logarithmically\cite{Kim02}
($\ell' \sim \log N$). However a real Web crawler would have to visit all the 
neighbors of a node to determine their degree, and moving to high-degree nodes
makes such a strategy useless for actual Web navigation ($\ell(\ell') \sim N$) given the
power-law degree sequence distribution\cite{Broder00,Kleinberg01,Adamic01}.

Kleinberg\cite{Kleinberg00} showed that if information about the geographic
location of nodes is available, then a greedy algorithm that always picks
the neighbor geographically closest to the target can yield $\ell \sim (\log N)^2$
if the link topology follows a $D$-dimensional lattice, with $2D$ local links
to lattice neighbors plus one long range connection per node, linking to another
node chosen with with probability $\Pr(r) \sim r^{-\alpha}$ where $r$ is 
the lattice distance between the two nodes and $\alpha$ is a constant clustering 
exponent. In this model the optimal path length is achieved for a critical
clustering exponent dependent on the dimensionality of the lattice ($\alpha = D$).

Kleinberg's model is inspired by social small-world networks where geographical knowledge 
exists, but in the Web hypertext the notion of geography is of little relevance and
the lattice model is unrealistic. However, a more relevant topological distance metric 
can be defined in the Web, namely the distance induced by the lexical similarity
between the textual content of pages. Let us define such a lexical distance
\begin{equation}
r(p_1,p_2) = \frac{1}{s(p_1,p_2)} - 1
\end{equation}
where $(p_1,p_2)$ is a pair of Web pages and 
$s$
%\begin{equation}
%s(p_1,p_2) = \frac{\sum_{k \in p_1 \cap p_2} w_{kp_1} w_{kp_2}}{\sqrt{\sum_{k \in p_1} w_{kp_1}^{2} \sum_{k \in p_2} w_{kp_2}^{2}}}
%\end{equation}
is the cosine similarity function traditionally used in information retrieval. 
%and $w_{kp}$ is some weight function for term $k$ in page $p$, e.g. term frequency.
The $r$ distance metric is a natural local cue readily available in the Web, with 
the target content specified by a query or topic of interest to the user. This metric
also does not suffer from the dimensionality bias that makes L-norms 
inappropriate in the sparse word vector space.

To investigate the relationship between the lexical topology induced by $r$ and
the link topology, I measured the frequency of linked pairs of pages 
as a function of the lexical distance, $\Pr(r(p_1,p_2)=\rho)$: 
\begin{equation}
\Pr(\rho|\lambda) \approx \frac{|(p_1,p_2) : r(p_1,p_2)=\rho \wedge l(p_1,p_2) > \lambda|}{|(p_1,p_2) : r(p_1,p_2)=\rho|}
\end{equation}
where the linkage between two pages was approximated by the overlap
\begin{equation}
l(p_1,p_2) = \frac{|U_{p_1} \cap U_{p_2}|}{|U_{p_1} \cup U_{p_2}|}
\end{equation}
and $U_p$ is the URL set representing $p$'s neighborhood (inlinks, outlinks, and 
$p$ itself). The threshold $\lambda$ models the ratio of local versus long range links, 
analogous to Kleinberg's dimensionality $D$. Larger $\lambda$ values imply
more sparse long range connections. 

\begin{figure}[hp]
	\centering
	\includegraphics{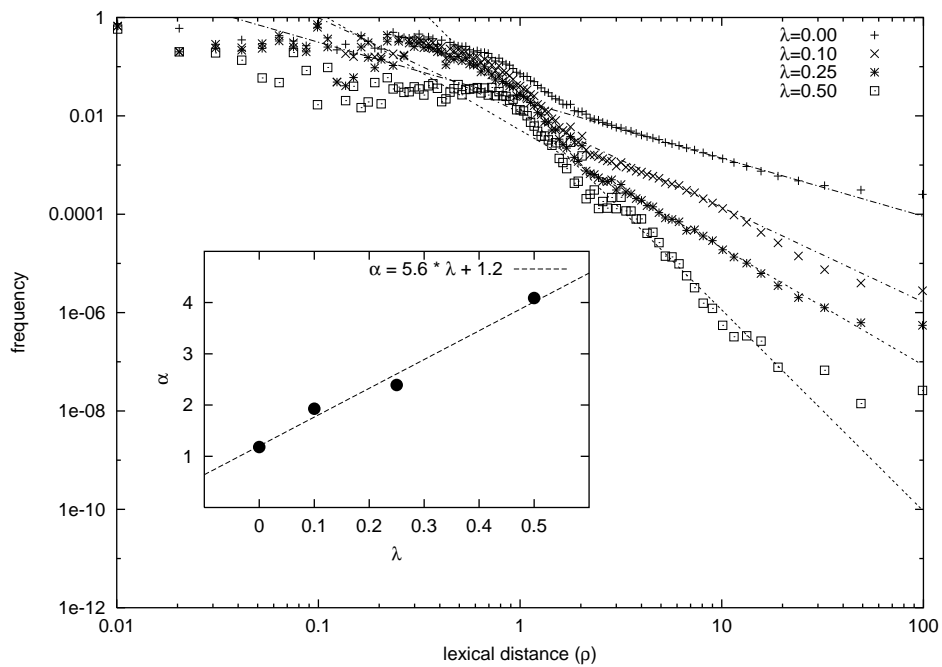}
	\caption{Linkage probability $\Pr(\rho)$ as a function of lexical distance $\rho$, 
	for various values of the linkage threshold $\lambda$. The frequency data is based on
	approximately $6 \times 10^8$ % 597,749,541 
	pairs of Web pages sampled from the Open Directory (\texttt{dmoz.org}).
	The least-squares fit of the tail of each distribution to the power law model 
	$\Pr(\rho) \sim \rho^{-\alpha}$ is also shown. The inset plots the relationship between the
	effective dimensionality induced by the linkage threshold $\lambda$ and the clustering
	exponent $\alpha$ of the power law tail.}
	\label{distance-linkage}
\end{figure}

Figure~\ref{distance-linkage} shows that long range links are indeed
distributed according to a power law 
$\Pr(\rho|\lambda) \sim \rho^{-\alpha(\lambda)}$, as in Kleinberg's 
model, but using lexical distance. A fit of the tails to the power law
model reveals that the clustering exponent $\alpha$ 
grows linearly with the linkage threshold $\lambda$ (inset).

Such a surprising result makes Kleinberg's analysis applicable to the Web. 
Crawlers that exploit textual cues may be able to navigate through short 
paths and locate unknown relevant pages. This is encouraging for
the community of crawling algorithm designers. The actual bound on the 
length of the path $\ell$ depends on whether the clustering exponent
is near a critical value; this will have to be determined by extending 
Kleinberg's analysis to the lexical topology of the Web.  

The observed relationship between link and lexical Web topology
has another interesting implication. One of the recent attempts
to explain the power law distribution of Web degree sequences is based
on a preferential attachment model in which new nodes are linked to existing
ones based on a critical mixture of linkage bias (attach to a 
node within a few links from most other nodes) and geographic bias
(attach to a node within a small Euclidean distance), where nodes
are given random coordinates in the unit square\cite{Papa02}. The
present result could lead to a more realistic interpretation 
whereby authors would link their new pages
to sites that are both popular and related in content, i.e., 
central in link space and nearby in lexical space. 

\begin{singlespace}

%\bibliography{menczer}

\end{singlespace}

\end{document}